\documentclass[a4paper,11pt]{article}
\usepackage{pos}
\usepackage{graphicx}
\usepackage{subcaption}
\usepackage{braket,amsmath,hyphenat,textcomp}
\usepackage[numbers,sort&compress]{natbib}
\newcommand{\beq}{\begin{equation}}
\newcommand{\eeq}{\end{equation}}

\title{Lattice calculation of transport coefficient $\hat{q}$ in pure gluon plasma and (2+1)-flavor QCD plasma}
 \ShortTitle{Lattice calculation of transport coefficient $\hat{q}$ in (2+1)-flavor QCD}

\author*[a]{Amit Kumar}
\author[a]{Abhijit Majumder}
\author[b]{Johannes Heinrich Weber}

\affiliation[a]{Department of Physics and Astronomy, Wayne State University, Detroit, MI 48201, USA\\
  }

\affiliation[b]{Department of Physics and Astronomy, Michigan State University, East Lansing, MI 48824, USA}

\emailAdd{kumar.amit@wayne.edu}
\emailAdd{majumder@wayne.edu}
\emailAdd{weberjo8@msu.edu}

\abstract{
The transport coefficient $\hat{q}$ is a leading coefficient that controls the modification of the hard parton traversing QGP, and hence, responsible for the suppression  of the high transverse momentum (transverse to the beam direction) charged-hadrons  in heavy-ion collisions.
In this article, we present the first unquenched lattice QCD calculation of $\hat{q}$. The calculation is carried out using (2+1)-flavor of quarks, using the highly improved staggered quark action (HISQ) and tree-level Symanzik improved gauge action. The calculation is performed in a wide range of temperatures, ranging from 200 MeV $<T<$ 800 MeV using MILC code package. We considered a leading-order process where a hard parton scatters off the glue field of a thermal QCD medium by exchanging a Glauber
gluon (whose transverse momentum is larger than its longitudinal components). The hard scale associated with the jet parton allows the coupling of the gluon to that parton to be treated in perturbation theory. The coupling of the gluon to the medium is treated
non-­perturbatively. This non-­perturbative part is expressed in terms of a non-­local (two­-point)  field-strength-­field-­strength  operator   product which  can  be  Taylor  expanded after analytic continuation to the deep ­Euclidean region. Such an expansion allows us to write $\hat{q}$ in terms of a series of local operators, which are suppressed by factors of the hard parton energy. The calculated $\hat{q}$ and its temperature dependence demonstrates reasonable agreement with the phenomenological extraction carried out by the JET collaboration.

}

\FullConference{%
  HardProbes2020\\
  1-6 June 2020\\
  Austin, Texas}

\begin{document}
\maketitle

\section{Introduction}

The suppression of high transverse momentum (high-$p_{\mathrm{T}}$) charged-hadrons and inclusive jets in relativistic heavy-ion collisions at RHIC and LHC is considered as an indicator of the presence of the strongly-coupled quark-gluon plasma (QGP). Among existing known coefficients characterizing the energy-loss of the hard parton traversing QGP, the transport coefficient $\hat{q}$ is the leading coefficient that controls the rate of medium-induced radiative energy loss of the hard parton inside QGP.
%
The coefficient $\hat{q}$ is defined as average squared transverse momentum broadening per unit length of the medium. 
Over the past years, several attempts have been made to compute $\hat{q}$ from first principles, each with its own assumptions and region of validity  \cite{Liu_2006,HTLcalculation,
Benzke:2012sz,  
Majumder:2012sh,
Panero:2013pla,    
Kumar:2019QGPPDF}.
In limit of high temperature, the hard-thermal-loop (HTL) perturbation theory predicts $\hat{q}$ to scale as a product of $T^3$ times $\mathrm{log}(E/T)$ \cite{HTLcalculation}.
A lattice gauge theory based formalism has also been proposed by the authors  of Ref. \cite{Majumder:2012sh,Benzke:2012sz,Panero:2013pla}.
A phenomenology based extraction of $\hat{q}$ has also been carried out by the JET \cite{Burke:2013yra} and JETSCAPE \cite{Soltz:2019aea} collaborations. Moreover, the work presented by the authors of Ref.  \cite{Kumar:2019QGPPDF} indicate that  $\hat{q}$ also possesses a dependence on the resolution scale of the hard parton.  

In these proceedings, we follow the methodology outlined in the article \cite{Majumder:2012sh} and compute $\hat{q}$ using lattice gauge theory.
We shall present our first estimates of the temperature dependence of $\hat{q}$ for the hard quark traversing the pure gluon plasma and 2+1 quark-flavor QCD plasma.

\section{Computing operators for pure gluon plasma and 2+1 flavor QCD plasma}
In this work, we briefly discuss the framework to compute $\hat{q}$ as outlined in Ref. \cite{Majumder:2012sh}. We consider the propagation of a hard quark carrying light-cone momentum 
$q=(\mu^2/2q^{-},q^{-}, 0_{\perp})\sim (\lambda^{2}, 1 , 0)q^{-}$ through a section of the plasma at temperature $T$, where, $ \lambda \ll 1 , q^{-} \gg  \Lambda_{\mathrm{QCD}} $, and $\mu$ is off-shellness of the hard quark.
We consider a leading order (LO) process, in which the
hard quark traveling along the negative $z$-direction
exchanges a transverse gluon with the plasma.  
In this frame $q^{0}>0$, $q_{z}<0$, and $q^0 \leq |q_z|$. Thus, the light-cone momentum of the quark $q^{+} = \frac{q^0+q_{z}}{\sqrt{2}} \leq 0$ and $q^{-} = \frac{q^0-q_{z}}{\sqrt{2}} \geq 0$.
We consider this process in the rest frame of the medium with the momentum of the exchanged gluon as $k=(k^{+}, k^{-}, k_{\perp}) \sim (\lambda^{2} , \lambda^{2},\lambda )q^{-}$.

Applying standard pQCD techniques, one express $\hat{q}$ for LO process as
\begin{equation}
\begin{split}
\hat{q} = \frac{8 \sqrt{2} \pi \alpha_{s} }{N_c}    \int \frac{ dy^{-} d^{2} y_{\perp} } {(2\pi)^3}  d^{2} k_{\perp}  e^{ -i  \frac{\vec{k}^{2}_{\perp}}{2q^{-}} y^{-} +i\vec{k}_{\perp}.\vec{y}_{\perp} } \sum _{n}  \bra{n} \frac{e^{-\beta E_{n}}}{Z} \mathrm{Tr[} F^{+ \perp _{\mu}}(0) F^{+}_{\perp_{\mu}}(y^{-},y_{\perp}) ] \ket{n},
\end{split}
\end{equation}
 where $F^{ \mu\nu} = t^{ a} F^{ a\mu\nu}$ is the gauge field strength, $\alpha_{s}$ is the strong coupling constant at the scale of the exchanged gluon, $\beta$ is the inverse temperature, $\ket{n}$ is a thermal state with energy $E_n$, $Z$ is the  partition function of the thermal medium, and $N_{c}$ is the number of colors. Computing the thermal expectation value of the operator $ F^{+ \perp _{\mu}}(0) F^{+}_{\perp_{\mu}}(y^{-},y_{\perp})$ on the lattice is challenging due to the light-cone separation of the two operators. However, using the method of dispersion relation, one can express $\hat{q}$ in terms of a series of local field-strength field-strength (FF) operators \cite{Majumder:2012sh,AKumarLattice}: 
\begin{equation}
\hat{q} = \frac{ 8\sqrt{2}\pi \alpha_{s}(\mu^{2})  }{N_c (T_{1} + T_{2})} \bra{M} \mathrm{Tr[} F^{+ \perp_{\mu}}(0) \sum^{\infty}_{n=0}   \left(   \frac{i\sqrt{2}D_{z} }{q^{-}}\right)^{n} F^{+}_{\perp_{\mu}}(0)  ]  \ket{M}_{(\mathrm{Thermal-Vacuum})},
\label{eq:qhatLatticeEquation}
\end{equation} 
where $T_{1}+T_{2} \approx 2T$, and $D_{z}$ is the covariant derivative along the $z$-direction.

The above expression of $\hat{q}$ is suitable for lattice calculation and can be used to extract $\hat{q}$ for both pure gluon plasma and quark-gluon plasma.
Note, in the above expression of $\hat{q}$, the higher-order terms in the series are suppressed by the hard scale $q^{-}$, and hence, computing the first few terms may be sufficient. It is also interesting to mention that a similar kind of operator products containing the covariant derivatives have been found by the author of Ref. \cite{Ji:2013dva} in the analysis of the parton distribution function on a  Euclidean space.
\begin{figure}[h]
  \centering
    \includegraphics[width=0.9\linewidth]{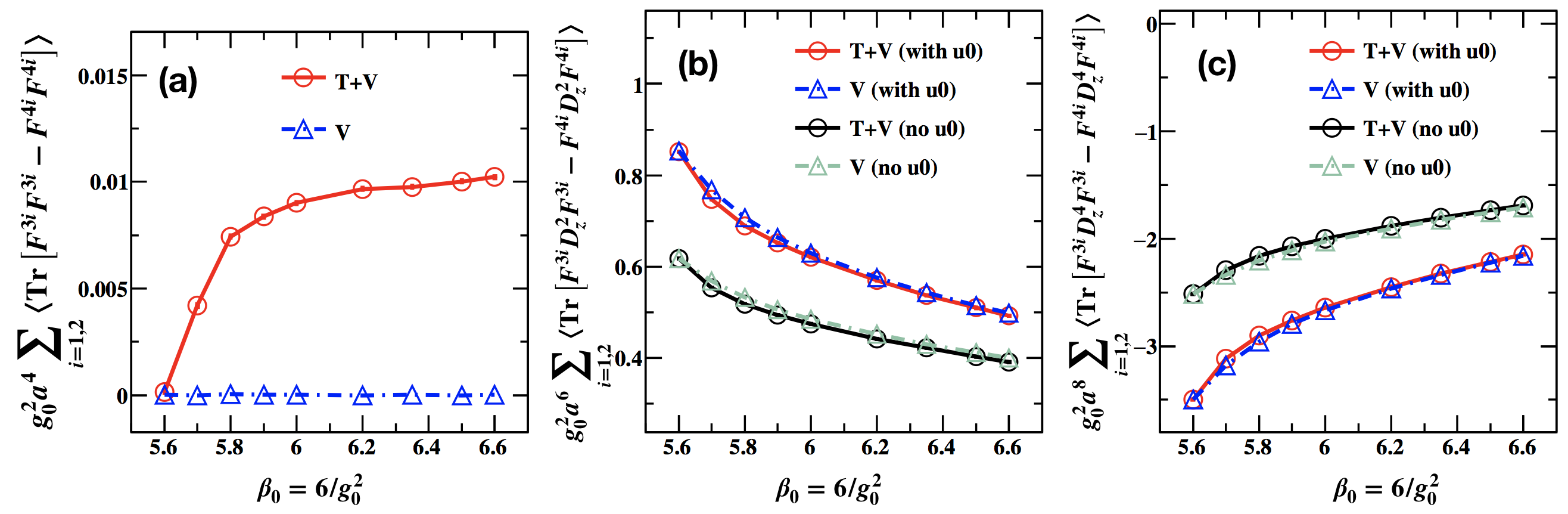}
  \caption{Thermal +Vacuum (T+V) and vacuum (V) expectation value of operators for pure $SU(3)$ lattice of size $n_{\tau}=4$, $n_{s}=16$. The label $u0$ represents tadpole improvements for the links in the covariant derivatives. (a) $g^{2}_{0}a^{4} \sum^{2}_{i=1}\mathrm{Tr}( F^{3i} F^{3i} - F^{4i} F^{4i}) $. (b)  $g^{2}_{0}a^{6} \sum^{2}_{i=1}\mathrm{Tr}( F^{3i} D^{2}_{z} F^{3i} - F^{4i} D^{2}_{z} F^{4i}) $.  (c)  $g^{2}_{0}a^{6} \sum^{2}_{i=1}\mathrm{Tr}( F^{3i} D^{4}_{z} F^{3i} - F^{4i} D^{4}_{z} F^{4i}) $.}
  \label{fig:BareOperatorNt4PureGauge}
\end{figure}
\begin{figure}[h]
  \centering
    \includegraphics[width=0.9\linewidth]{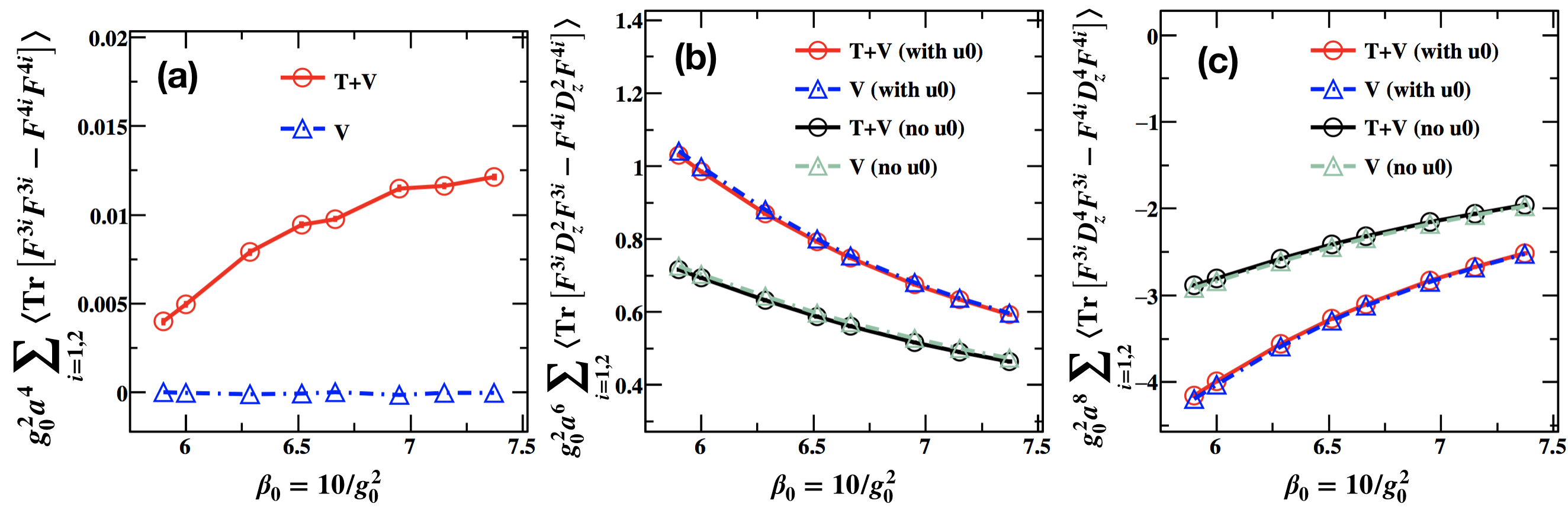}
  \caption{Same as Fig. \ref{fig:BareOperatorNt4PureGauge}, except the operators are evaluated on (2+1)-flavor unquenched lattice. }
  \label{fig:BareOperatorNt4FullQCD}
\end{figure}

In our first attempt, we consider the following three non-zero operators in $\hat{q}$ series:
 $\sum^{2}_{i=1}\mathrm{Tr}( F^{3i} F^{3i} - F^{4i} F^{4i}) $,   $\sum^{2}_{i=1}\mathrm{Tr}( F^{3i} D^{2}_{z} F^{3i} - F^{4i} D^{2}_{z} F^{4i}) $ and   $\sum^{2}_{i=1}\mathrm{Tr}( F^{3i} D^{4}_{z} F^{3i} - F^{4i} D^{4}_{z} F^{4i}) $.
To compute the field strength tensor on the lattice, we employed a clover-leaf discretization, given as
$F_{\mu\nu}(x) = \left[ Q_{\mu\nu}(x) - Q_{\mu\nu}^{\dagger}(x) - \frac{1}{3}\mathrm{Tr}(Q_{\mu\nu}(x) - Q_{\mu\nu}^{\dagger}(x))  \right] / (8ig_{0}a^{2}_{L}),
$
where $Q_{\mu\nu}(x)= U_{\mu,\nu}(x) + U_{-\mu,\nu}(x) + U_{-\mu,-\nu}(x)+ U_{\mu,-\nu}(x)$ represents the sum over four plaquettes ($U_{\mu,\nu}$) around the site $x$ in the $\mu$-$\nu$ plane (anti-clockwise direction), $g_{0}$ is the bare lattice coupling and $a_{L}$ is the lattice spacing. 
%
%
In the calculation of the equation of state (EOS), the trace anomaly $(\epsilon -3p)/T^4$ is computed by evaluating the vacuum subtracted expectation value of the gluon Lagrangian density $(-1/4)F_{\mu\nu}F^{\mu\nu}$, where the lattice beta function, $R_{\beta_{0}}=-ad\beta_{0}/da$, appears as a multiplicative renormalization factor \cite{Bazavov_2014}. Since, the FF operators in our case are similar to operators in trace anamoly, the FF operators $\sum \limits_{i=1,2} (F^{2}_{3i}-F^{2}_{4i})$ must have $R_{\beta_{0}}$ as the multiplicative renormalization factor. 

The gauge field configurations for both quenched and unquenched plasma are generated using the public version of Multiple Instruction $\&$ multiple data (MIMD) Lattice Computation (MILC) code package \cite{MILCcodePackage}.
The thermal configurations are generated with the aspect ratio $n_{s}/n_{\tau}=4$, whereas the corresponding vacuum configurations are generated with $n_{\tau}=n_{s}$. 
For pure $SU(3)$ plasma, we employed Wilson's pure SU(3) gauge action, where the input parameter $\beta_{0}$ is given as $\beta_{0}=6/g^{2}_{0}$.
The gauge configurations with (2+1)-flavors of quarks are generated  using the highly improved staggered quark action (HISQ) and tree-level Symanzik improved gauge action \cite{Bazavov_2014,Bazavov_2018}.
The calculations have been done by taking a statistical average over 10000 gauge configurations generated using the Rational Hybrid Monte-Carlo  algorithm. For the unquenched case, we employed tuned input parameters (bare coupling, quark masses) published in Ref. \cite{Bazavov_2014,Bazavov_2018} by the HotQCD and TUMQCD Collaboration. The strange quark mass  $m_{s}$ was set to the physical value with the degenerate light quark masses $m_{u,d}=m_{s}/20$; in the continuum limit, this corresponds to a pion mass of about 160 MeV. The input parameter $\beta_{0}$ is related to the bare gauge coupling as $\beta_{0}=10/g^{2}_{0}$.


\begin{figure}[h!]
  \centering
    \includegraphics[width=0.7\linewidth]{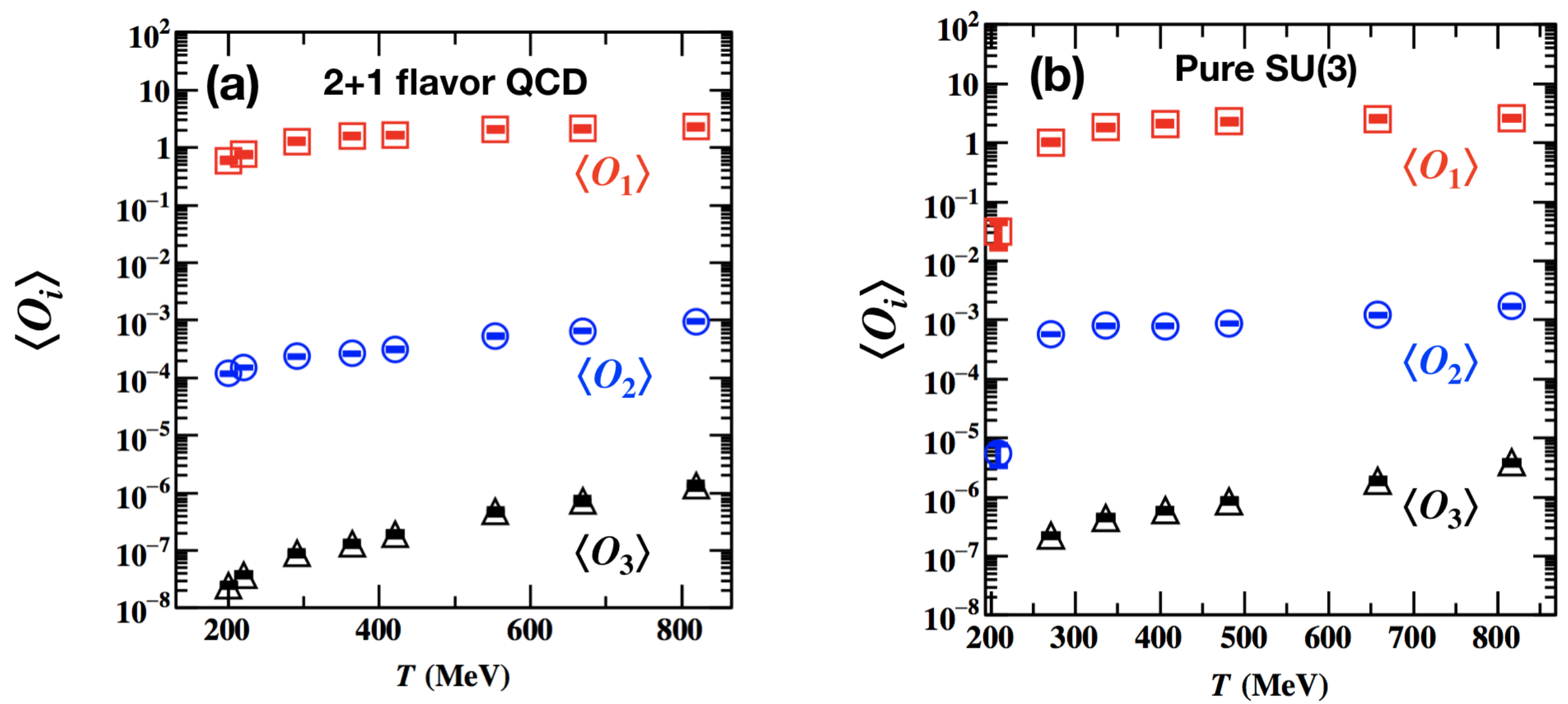}
  \caption{Temperature dependence of vacuum subtracted FF correlators for lattice size $n_{\tau}$=4, $n_{s}$=16. $ \left \langle O_{1} \right \rangle=\mathrm{Tr}[F^{3i}F^{3i} - F^{4i}F^{4i}]/T^{4}$, $\left \langle O_{2} \right \rangle=\mathrm{Tr}[F^{3i}D^2_{z}F^{3i} - F^{4i}D^2_{z}F^{4i}]/(T^{4}(q^{-})^2)$, and $ \left \langle O_{3} \right \rangle =\mathrm{Tr}[F^{3i}D^4_{z}F^{3i} - F^{4i}D^4_{z}F^{4i}]/(T^{4}(q^{-})^4)$.   We set $q^{-}=100$ GeV. (a) 2+1 flavor QCD. (b) Pure SU(3) gauge.  }
  \label{fig:TemperatureDependenceOfLatticeOperators}
\end{figure}
Fig. \ref{fig:BareOperatorNt4PureGauge} and \ref{fig:BareOperatorNt4FullQCD} shows expectation value of FF operators on pure SU(3) quenched and 2+1 flavor unquenched lattices, respectively. The vacuum (V) expectation value of $\mathrm{Tr}[F^{3i}F^{3i} - F^{4i}F^{4i}]$ is negligible compared to its thermal + vacuum (T+V) expectation value. For operators with $D_{z}$ derivatives, T+V and V results have similar magnitude and sign; we also note T+V and V results enhance by the same factor as one adds tadpole  factor ($u_{0}$) for the links in the $D_{z}$ derivative.

Fig. \ref{fig:TemperatureDependenceOfLatticeOperators}(a) and  \ref{fig:TemperatureDependenceOfLatticeOperators}(b) displays vacuum subtracted expectation value of FF operators on 2+1 flavor unquenched and pure SU(3) quenched lattices, respectively. The operator $\mathrm{Tr}[F^{3i}F^{3i} - F^{4i}F^{4i}]/T^{4}$ is dominant compared to FF operators with $2^{\mathrm{nd}}$-order $D_{z}$ derivative  and $4^{\mathrm{th}}$-order $D_{z}$ derivative.
For the full QCD case, our results show a smoother increase in the crossover region.

\section{Results and Discussions}
\begin{figure}[h!]
  \centering
    \includegraphics[width=0.5\linewidth]{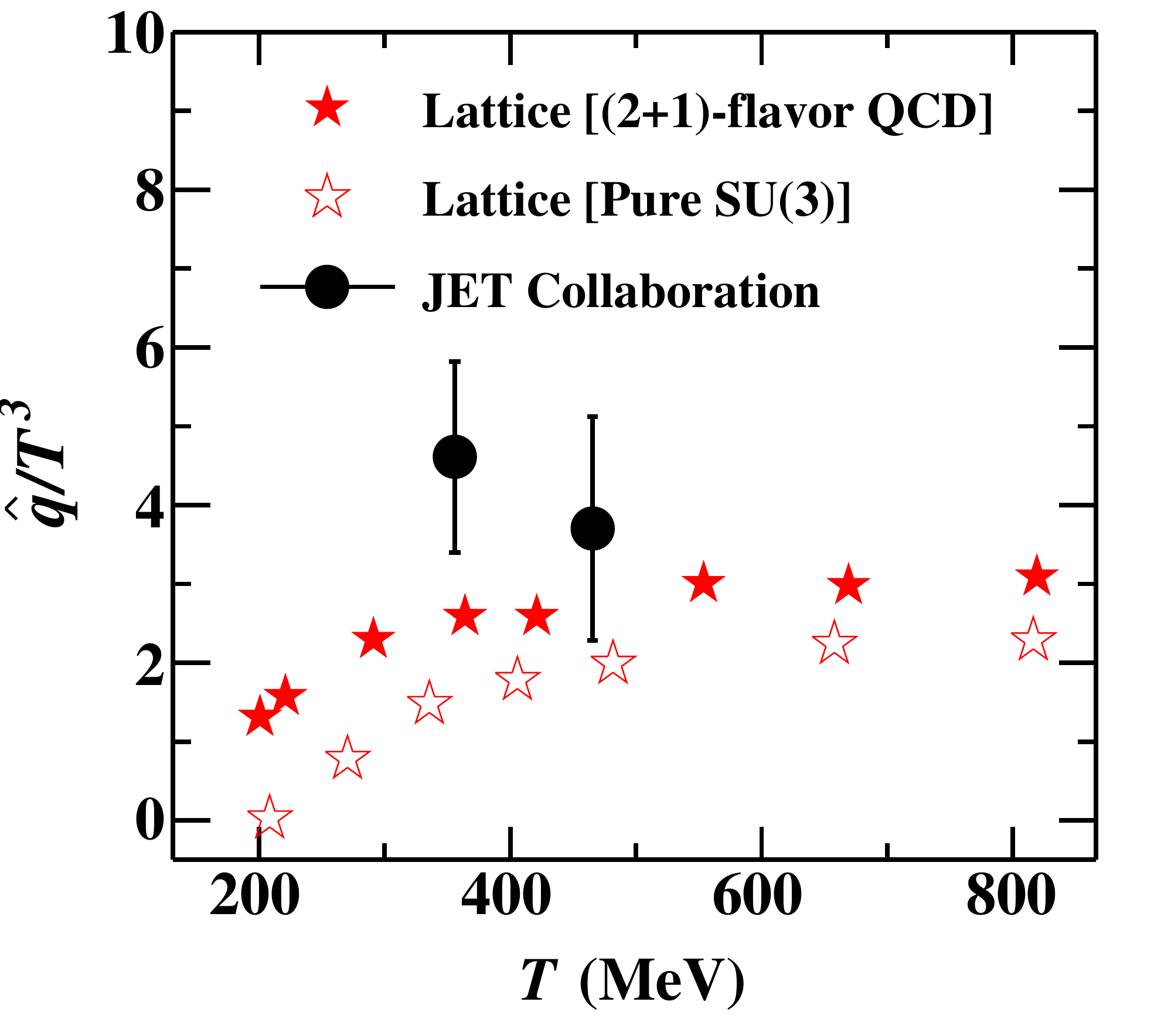}
  \caption{$\hat{q}/T^3$ as a function of temperature  for a pure gluonic plasma and (2+1)-flavor QCD plasma computed on lattice of size $n_{\tau}=4$, $n_{s}=16$.  A comparison with JET collaboration \cite{Burke:2013yra} result is also shown.}
  \label{fig:qhatT3WrtT}
\end{figure}

The light-cone momentum (energy) $q^-$ of the hard quark was set to 100 GeV. 
We evaluate $\alpha_{s}$ in Eq. \ref{eq:qhatLatticeEquation} at the scale $\mu^{2}=(\pi /a)^{2}= (\pi Tn_{\tau})^2$. This scale is the same as the scale at which our FF correlators are computed on the lattice due to the momentum cutoff. 
Fig. \ref{fig:qhatT3WrtT} shows temperature dependence of $\hat{q}/T^3$ for pure gluon plasma and (2+1)-flavor QCD plasma computed on the lattice of size $n_{\tau}=4$ and $n_{s}=16$. At high temperatures,  $\hat{q}$ scales with $T^3$; we obtain $\hat{q}/T^3 \sim $1.5-2.5 for quenched lattices and $\hat{q}/T^3 \sim $2.5-3.5 for unquenched lattices. 
Fig. \ref{fig:qhatT3WrtT} also shows our lattice results are consistent with phenomenology based extraction carried out by the
JET collaboration \cite{Burke:2013yra} (within $2\sigma$ uncertainty). 


In these proceedings, we computed $\hat{q}$ for the hard quark traversing  pure gluonic and (2+1)-flavor QCD plasma for lattice size $n_{\tau}=4$ and $n_{s}=16$. The calculation for finer lattice sizes is ongoing.

\textbf{Acknowledgments:}
This work was supported in part by the National Science Foundation under the grant No. ACI-1550300 within the JETSCAPE collaboration, and in part by US department of energy, office of science, office of nuclear physics under grant No. DE-SC0013460.

\end{document}